\documentclass[sigplan,10pt]{acmart}

\usepackage{subcaption}
\usepackage{algorithm}
\usepackage{algpseudocode}
\usepackage{pifont}
\usepackage{makecell}
\usepackage{threeparttable}

\setcopyright{none}
\AtBeginDocument{%
  }

\acmISBN{}
\acmDOI{}
\acmYear{}
\copyrightyear{}
\acmPrice{}

\newcommand{\sysname}{SpecGen}

\renewcommand\footnotetextcopyrightpermission[1]{}
\renewcommand\footnotetextcopyrightpermission[1]{}
\makeatletter
\AtBeginDocument{%
\fancypagestyle{standardpagestyle}{%
  \fancyhf{}%
  \renewcommand{\headrulewidth}{\z@}%
  \renewcommand{\footrulewidth}{\z@}%
  \fancyfoot[C]{\if@ACM@printfolios\footnotesize\thepage\fi}
  \fancyhead[LE]{\ACM@linecountL}
  \fancyhead[LO]{\ACM@linecountL}%
  \fancyhead[RE]{\ACM@linecountR}%
  \fancyhead[RO]{\ACM@linecountR}%
}%
\pagestyle{standardpagestyle}%
}

\makeatother

\begin{document}

\title[SpecGen: Accelerating Agentic Kernel Optimization with Speculative Generation]{%
  {\fontsize{18}{22}\selectfont SpecGen: Accelerating Agentic Kernel Optimization with Speculative Generation}%
}

\author{Jihu Guo}
\authornote{These authors contributed equally to this research.}
\email{24110240025@m.fudan.edu.cn}
\affiliation{%
  \institution{Fudan University \& Shanghai AI Lab}
  \country{China}
}

\author{Sitian Lu}
\authornotemark[1]
\email{sitianlu@sjtu.edu.cn}
\affiliation{%
  \institution{Shanghai Jiao Tong University \& Shanghai AI Lab}
  \country{China}
}

\author{Tenghui Ma}
\email{24110240061@m.fudan.edu.cn}
\affiliation{%
  \institution{Fudan University \& Shanghai AI Lab}
  \country{China}
}

\author{Wei Gao}
\authornote{Corresponding author.}
\email{csgaowei@ust.hk}
\affiliation{%
  \institution{Hong Kong University of Science and Technology}
  \country{China}
}

\author{Zhisheng Ye}
\email{yezhisheng@pku.edu.cn}
\affiliation{%
  \institution{Independent Researcher}
  \country{China}
}

\author{Xingcheng Zhang}
\email{zhangxingcheng@pjlab.org.cn}
\affiliation{%
  \institution{Shanghai AI Lab}
  \country{China}
}

\author{Dahua Lin}
\email{dhlin@ie.cuhk.edu.hk}
\affiliation{%
  \institution{The Chinese University of Hong Kong \& Sensetime Research}
  \country{China}
}

\begin{abstract}

Agentic kernel optimization automates manual GPU kernel tuning via
iterative generation, validation, and profiling with reasoning LLMs, casting the optimization task as
feedback-guided search. However,
our workload characterization reveals three system-level inefficiencies that limit search efficiency: 
(1) long generation latency due to LLM reasoning,
(2) insufficient profiling feedback, and
(3) underutilized validation/profiling resources.
Our key insight is that the ongoing reasoning generation exposes a window for
producing additional candidate kernels before it completes, allowing the system
to terminate reasoning early once a satisfactory kernel appears.

We present \sysname{}, an agentic kernel optimization system with \emph{speculative generation}.
First, \sysname{} forks 
non-reasoning generations at well-chosen trigger points in the
reasoning trace to yield kernels, increasing the candidate kernel count per iteration. 
These kernels are validated and profiled in parallel with the
ongoing reasoning, increasing
profiling feedback, and keeping resources busy during generation.
When a kernel meets the termination criterion, \sysname{} terminates the reasoning
generation early to reduce the generation latency.
Second, \sysname{} dynamically reallocates validation and profiling GPU pools based
on the arrival rate and prioritizes requests to reduce profiling feedback latency 
under bursty speculative generation load. 
Furthermore, \sysname{} utilizes spare memory of the validation/profiling GPUs as remote 
KV cache storage to eliminate prefix recomputation of speculative generations under 
limited memory budget.
Experiments with two reasoning LLMs on H200 show that
\sysname{} reduces 1.68--1.82$\times$ end-to-end time over three baseline
systems, while producing 1.58--1.98$\times$ profiling feedback, increasing
resource utilization from 4.2--17.6\% to 88.2--96.1\%, and improving 1.24--1.91$\times$ kernel 
speedup under a fixed time and token budget.

\end{abstract}

\maketitle

\section{Introduction}
\label{sec:intro}

GPU kernels underpin every layer of modern LLM training and inference.
Even a few percent improvement to attention, GEMM, or normalization kernels translates directly into reduced infrastructure cost~\cite{kevin32b,cudaagent,cudaforge}.
However, kernel optimization is time and labor intensive, costing experts weeks of tuning kernel performance on specific GPUs~\cite{flashattention1,flashattention2,flashattention3}.
Agentic kernel optimization automates the manual kernel tuning procedure with reasoning LLMs. 
It casts the kernel optimization process as a feedback-guided iterative search over a large kernel design space~\cite{K-Search,Kernel-Smith,AlphaEvolve,cudaforge}.
Each iteration performs three phases in sequence:
\textbf{(1)} \emph{generation}: calling reasoning LLMs to generate candidate kernels;
\textbf{(2)} \emph{validation}: compiling and running these kernels against a reference kernel to check correctness and a first-cut speedup;
and \textbf{(3)} \emph{profiling}: measuring detailed performance metrics and adding them to the LLM's context to inform subsequent generations.
The objective of agentic kernel optimization is to \emph{maximize kernel performance under a fixed time or token budget}.

However, the large kernel design space makes this search time- and resource-costly, with
a kernel optimization task often taking tens of hours or even days~\cite{cudaforge,AlphaEvolve,kernelagent2025}.
To understand where the budget goes, we conduct a comprehensive workload characterization (see \S\ref{sec:mot}) of agentic kernel optimization across
10 \texttt{KernelBench}~\cite{kernelbench} tasks on an H200 testbed with two state-of-the-art open-source reasoning LLMs, 
GLM-5.1~\cite{glm5} and DeepSeek-V4-Pro~\cite{deepseekv4}.
Our workload characterization identifies three system-level inefficiencies that limit search efficiency: 
\textbf{(C1)} long generation latency: the generation phase accounts for 70--99\% of iteration time, directly capping the iteration count under fixed time budget;
\textbf{(C2)} insufficient profiling feedback: about 59--64\% of iterations consume GPU hours without producing profiling feedback to drive the next iteration's optimization; 
and \textbf{(C3)} underutilized validation and profiling resources: validation
and profiling GPUs achieve only 4.2--17.6\% utilization.
They remain idle during the generation phase because candidate kernels emerge
only after generation completes, and many candidates fail validation before profiling.

To address these inefficiencies, prior systems adopt different harness engineering
strategies, such as best-of-$K$~\cite{cudaforge,K-Search}, multi-agent~\cite{kernelagent2025}, 
and evolutionary algorithms~\cite{AlphaEvolve,Kernel-Smith}.
Despite their strategies, these systems share a common pattern. 
They dispatch multiple reasoning generations per iteration to increase the
candidate count and produce more profiling feedback, but this roughly multiplies
token cost by $K$. 
A strawman alternative is to dispatch non-reasoning generations to reduce token cost. 
However, non-reasoning generations rarely produce valid or performant kernels (see Table~\ref{tab:nonreasoning_speedups_w_wo_shared_prefix_cache}).
Moreover, prior systems 
do not reduce generation latency since candidate kernels become actionable only after the 
generation finishes, and validation/profiling resources remain idle during the reasoning 
process of the generation phase.

Our key insight is that the reasoning process exposes a window for producing
additional candidate kernels before the reasoning generation completes
(\S\ref{subsec:design_insight}).
Since non-reasoning generations are faster than reasoning generations, the system can dispatch
multiple non-reasoning generations in parallel with the ongoing reasoning generation to 
increase the candidate kernel count.
To improve the kernel performance of non-reasoning generations, we condition non-reasoning
generations on the prefix of the reasoning output.
With this conditioning, the non-reasoning generations can produce valid and performant kernels before the
reasoning generation completes (see Table~\ref{tab:nonreasoning_speedups_w_wo_shared_prefix_cache}).
Our further experiments show that prefix conditioning can even generate kernels with higher speedups than 
the average speedup of previously generated kernels (see Figure~\ref{fig:spec_reach95_token_cdf}).
This provides three benefits:
\textbf{(1)} shorter generation latency by terminating the reasoning
process early when a satisfactory candidate emerges (\textbf{C1});
\textbf{(2)} more profiling feedback through generating more candidate kernels per
iteration (\textbf{C2});
\textbf{(3)} higher validation/profiling utilization by keeping validation/profiling
resources busy during generation (\textbf{C3}).
We empirically show that a pragmatic termination criterion reduces generation
latency without compromising kernel performance
(see \S\ref{subsec:eval_termination_analysis}).
Building on this insight, we introduce \emph{speculative generation}, which
forks non-reasoning generations conditioned on the reasoning prefix.
Because each fork shares the cached reasoning prefix, speculative generation
incurs low token overhead (see \S\ref{subsec:eval_token_consumption}).
Moreover, speculative generation operates at the iteration level and is
orthogonal to token-level speculative decoding~\cite{specdec,medusa}.
The two techniques can be combined to further improve system efficiency.

However, realizing speculative generation introduces two system-level challenges.
\textbf{First,} \emph{how} to fork and \emph{when} to terminate:
the system must identify useful conditioning context from a noisy reasoning
trace and decide how many speculative generations to launch.
Too few forks leave validation/profiling resources idle, while too many overload
the queues and delay profiling feedback.
Terminating too early sacrifices kernel performance, while terminating too late saves negligible generation time.
\textbf{Second,} \emph{how} to manage bursty speculative-generation load.
The legacy static ``one GPU per kernel'' partitioning cannot handle this burstiness.
It leaves some GPUs idle while others queue, so resource allocation must adapt as bursts arrive.
The system also needs request prioritization to bound profiling feedback latency
and memory management to support bursty speculative generations in local LLM
deployments.

We present \sysname{}, an agentic kernel-optimization system that wraps a user-specified LLM, prompt, and search algorithm with two cooperating components.
\textbf{(1) SpecController} monitors the reasoning model's output stream and parses it to detect trigger signals, such as kernel-design decisions, fenced code blocks, and closed kernel bodies.
On a clean trigger, SpecController concatenates the iteration's prompt with the reasoning prefix and dispatches $K$ non-reasoning speculative generation requests.
The number of forks adapts to the available validation/profiling resource capacity, exposing enough candidates to keep GPUs busy without overloading the queues.
Once a speculatively generated kernel exceeds the average speedup of previously generated kernels, \sysname{} terminates the reasoning generation early to reduce the generation latency.
This threshold avoids terminating on weak speculative kernels while still allowing frequent early terminations.
\textbf{(2) ElasticScheduler} manages one elastic GPU pool, dynamically split between validation and profiling.
Each request takes any free GPU in its pool, so no GPU idles while another queues up.
Between iterations, ElasticScheduler reallocates the two pools based on the arrival rates of the previous iteration.
Within an iteration, the validation queue is served last-arrival-first.
Later candidates carry more reasoning prefix and are more likely to validate.
The profiling queue is served FIFO, so any validated kernel returns feedback fast and SpecController decides on the freshest signal.
We also observe that validation/profiling GPUs retain substantial unused memory
even under bursty speculative load.
We therefore repurpose this spare memory as remote storage for the
reasoning prefix cache, eliminating prefix recomputation of speculative generations.

Across \texttt{KernelBench} Level~1/2/3 tasks with two reasoning LLMs on H200,
\sysname{} consistently improves the efficiency of agentic kernel optimization
against CudaForge~\cite{cudaforge}, AlphaEvolve~\cite{AlphaEvolve}, and KernelAgent~\cite{kernelagent2025}.
Compared with these baselines, \sysname{} reduces end-to-end execution time by
1.68--1.82$\times$ and increases profiling feedback by 1.58--1.98$\times$.
It also improves final kernel speedup by 1.24--1.91$\times$, showing that
shorter E2E time does not come at the cost of kernel performance.
Finally, \sysname{} increases validation/profiling utilization from
4.2--17.6\% to 88.2--96.1\%.

This paper makes the following contributions:
\begin{itemize}
    \item \textbf{Characterization.} We conduct a comprehensive workload characterization of agentic kernel optimization and identify three system-level inefficiencies.
    \item \textbf{Insight.} We empirically establish that the reasoning process exposes a window for
    speculative generations conditioned on the reasoning prefix.
    \item \textbf{System.} We implement \sysname{}, comprising a SpecController and an ElasticScheduler that realize the above insight to address the three inefficiencies without sacrificing kernel performance.
    \item \textbf{Evaluation.} We evaluate \sysname{} on 20 \texttt{KernelBench} Level~1/2/3 tasks across two reasoning LLMs on H200 against three state-of-the-art agentic kernel optimization systems, with detailed results in \S\ref{sec:eval}.
\end{itemize}

\section{Background}\label{sec:bg}

\begin{figure}[t]
    \centering
    \includegraphics[width=\linewidth]{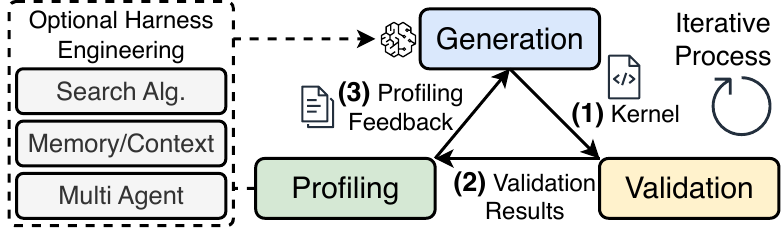}
    \vspace{-20pt}
    \caption{The three phases of agentic kernel optimization iteration: generation, validation, and profiling.}
    \label{fig:agentic_kernel_optimization}
    \vspace{-20pt}
\end{figure}

GPU kernel optimization is an iterative and verifiable task. Realizing a competitive kernel typically requires experts weeks per operator on a given GPU~\cite{Kernel-Smith,AlphaEvolve,cudal1,cudaagent}.

\noindent\textbf{Agentic Kernel Optimization} iteratively performs three phases as shown in Figure~\ref{fig:agentic_kernel_optimization} to generate and optimize kernels.
{\textbf{(1)} Generation} invokes the LLM on the accumulated context to produce a candidate kernel (e.g., CUDA/C++~\cite{kernelbench} or Triton~\cite{TritonBench}).
{\textbf{(2)} Validation} checks the candidate kernel for compilation, execution, and output correctness against a reference~\cite{kernelbench,TritonBench}. Any failure (compile error, runtime error, or numerical mismatch) is merged into the LLM's context to guide the next iteration's correction. Only kernels passing all three checks proceed to profiling.
{\textbf{(3)} Profiling} measures the candidate kernel's detailed performance metrics using NVIDIA Nsight Compute (\texttt{NCU}). The resulting metrics are merged into the LLM's context as feedback to guide subsequent generations.
Both validation and profiling require exclusive GPU access for measurement accuracy, since GPU sharing would distort both the speedup measured against the reference kernel and the profiled performance metrics.
Due to this exclusivity, existing frameworks statically dedicate one GPU per kernel to run its validation and profiling~\cite{cudaforge,TritonForge}, a partitioning we refer to as ``one GPU per kernel''.

\noindent\textbf{Harness Engineering Approaches} are also explored for agentic kernel optimization, including multi-agent collaboration~\cite{cudaforge,KernelSkill,Astra} and memory/context management~\cite{STARK,CuTeGen,K-Search}, evolutionary search~\cite{AlphaEvolve,openevolve,kernelagent2025,Kernel-Smith,KernelEvolve}, and verifier-guided search~\cite{ImprovingEfficiencyOfGPUKernelOptimizationAgents,AutoKernel,SwizzlePerf} to improve the kernel performance (Figure~\ref{fig:agentic_kernel_optimization} left). These works differ in their agentic optimization algorithms but still rely on the same three-phase iteration: invoking the LLM to generate kernels, validating their correctness against a reference, and profiling them for performance metrics.
Despite the rapidly evolving algorithmic frontier, the system-level efficiency of existing harness engineering approaches has received little attention. To analyze it, we characterize the agentic kernel optimization workload in \S\ref{sec:mot}.
\section{Workload Characterization}\label{sec:mot}

We characterize the workload of agentic kernel optimization on top of the iterative refinement framework from \texttt{KernelBench}~\cite{kernelbench}, an efficient agentic kernel optimization framework for improving kernel performance.
We run two reasoning LLMs, GLM-5.1~\cite{glm5} and DeepSeek-V4-Pro~\cite{deepseekv4}, on 10 \texttt{KernelBench}~\cite{kernelbench} kernel optimization tasks widely used in LLM training and inference (Table~\ref{tab:tasks}, 100 iterations per LLM per kernel), with H200 GPUs dedicated per kernel for validation and profiling.
The resulting traces reveal three system-level inefficiencies, which we analyze in \S\ref{subsec:gen_dominates}--\S\ref{subsec:gpu_idle}.

\begin{table}[t]
    \centering
    \caption{Short names (T1--T10) of the ten \texttt{KernelBench} kernels widely adopted in LLM training and inference.}
    \vspace{-10pt}
    \label{tab:tasks}
    \small
    \setlength{\tabcolsep}{5pt}
    \renewcommand{\arraystretch}{0.90}
    \begin{tabular}{@{}llll@{}}
    \toprule
    \textbf{Abbr.} & \textbf{Task} & \textbf{Abbr.} & \textbf{Task} \\
    \midrule
    T1 & HingeLoss        & T6  & Upper-tri. Matmul \\
    T2 & 3D tensor Matmul & T7  & Lower-tri. Matmul \\
    T3 & 4D tensor Matmul & T8  & $A^{\top}B$ Matmul \\
    T4 & Diagonal Matmul  & T9  & $AB^{\top}$ Matmul \\
    T5 & Symmetric Matmul & T10 & $A^{\top}B^{\top}$ Matmul \\
    \bottomrule
    \end{tabular}
    \vspace{-10pt}
\end{table}

\begin{figure}
    \centering
    \includegraphics[width=\linewidth]{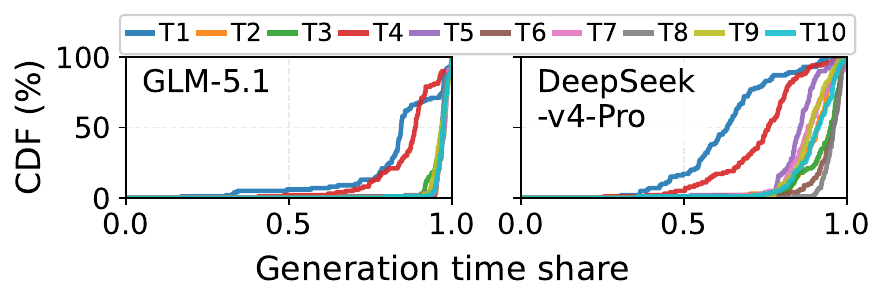}
    \vspace{-25pt}
    \caption{Per-iteration generation time share across ten \texttt{KernelBench} kernel optimization tasks per model.}
    \vspace{-15pt}
    \label{fig:generation_time_share_cdf}
\end{figure}

\subsection{Long Generation Latency}\label{subsec:gen_dominates}

We define the per-iteration \emph{generation time share} as the fraction of iteration time spent in the generation phase.
Figure~\ref{fig:generation_time_share_cdf} reports its empirical CDF across both reasoning models and the ten \texttt{KernelBench} tasks.
Under GLM-5.1, the task-level 75th-percentile (P75) of this share falls between 92--99\%, with most matmul variants clustered near 98\%. For DeepSeek-V4-Pro, P75 spans 70\%--98\%, with HingeLoss (T1) as a low outlier at 70\% and the remaining matmul workloads at P75 between 81\% and 98\%.
Across both models and the vast majority of iterations, LLM generation thus accounts for the bulk of each iteration's wall-clock, leaving validation and profiling as second-order time costs on the critical path.
In absolute terms, generation averages 706.9\,s and 522.6\,s per iteration on the two models, an order of magnitude above validation (22.9\,s and 59.0\,s) and profiling (26.5\,s and 26.6\,s).

\noindent\textbf{Reason.}
Kernel optimization is an intrinsically multi-axis problem. Each iteration must reconcile hardware constraints,
I/O and precision requirements, and evolving compute-versus-memory bottlenecks against the previous iterations' profiling
feedback. Models therefore need to reason longer to produce a kernel, leading to longer generation latency.

\noindent\textbf{Implication 1.} Long generation latency limits the iteration count under any fixed time budget, motivating a scheme to reduce the generation latency.

\begin{figure}
    \centering
    \includegraphics[width=\linewidth]{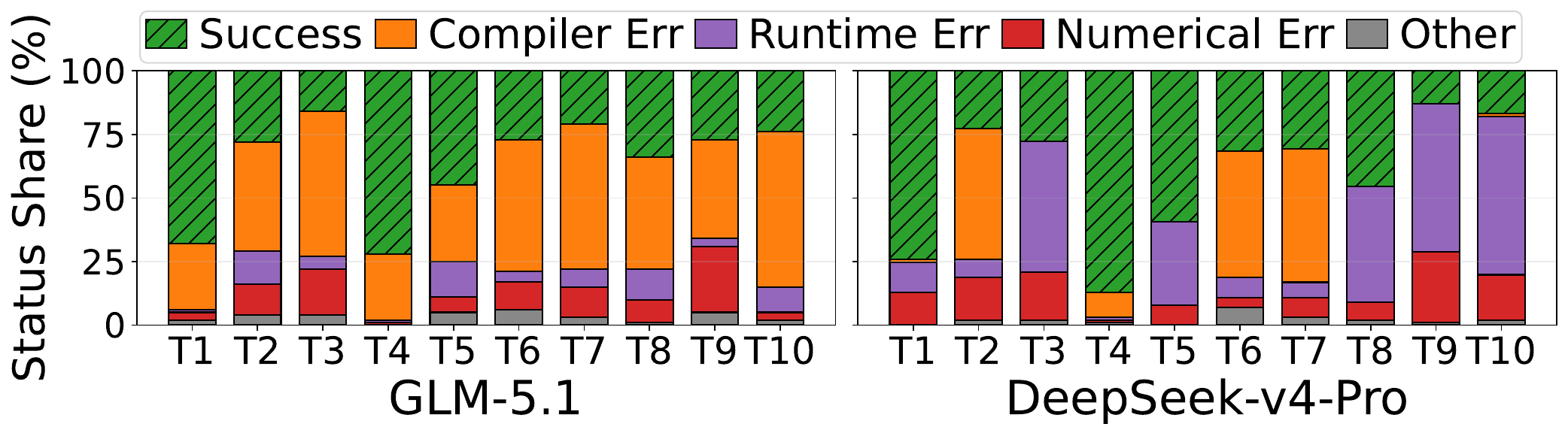}
    \vspace{-20pt}
    \caption{Distribution of iteration status across 100 iterations per kernel with two models. An iteration is \emph{Success} iff its emitted kernel compiles, runs, and matches the reference.}
    \label{fig:iteration_status_statistics}
    \vspace{-15pt}
\end{figure}

\begin{figure*}
    \begin{subfigure}[t]{0.49\linewidth}
        \centering
        \includegraphics[width=\linewidth]{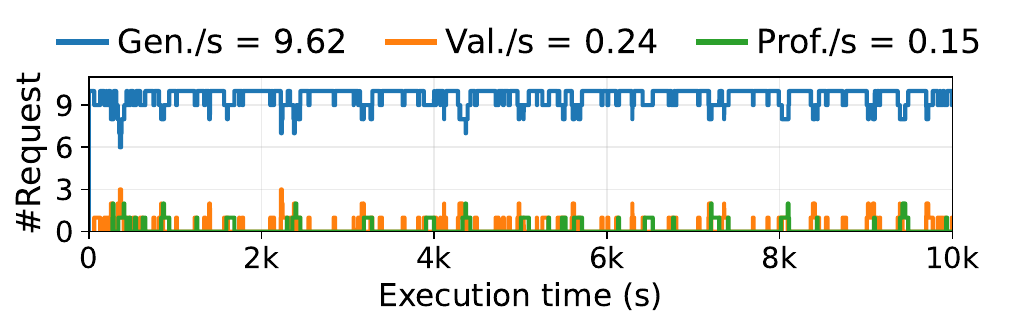}
        \vspace{-20pt}
        \caption{GLM-5.1.}
        \label{fig:phase_request_timelines_GLM_5_1}
    \end{subfigure}
    \begin{subfigure}[t]{0.49\linewidth}
        \centering
        \includegraphics[width=\linewidth]{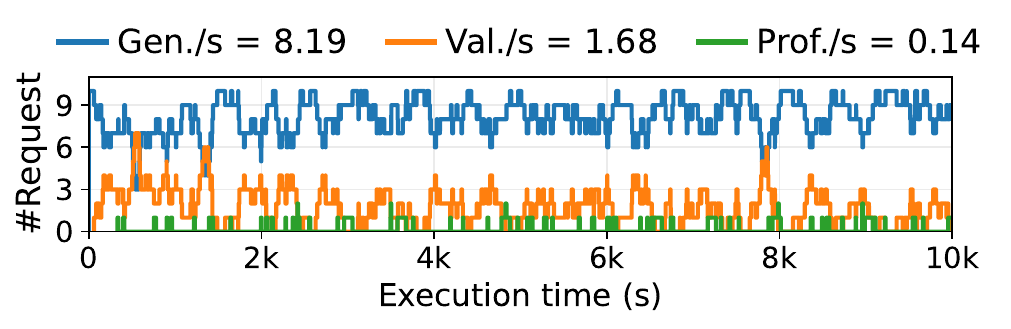}
        \vspace{-20pt}
        \caption{DeepSeek-V4-Pro.}
        \label{fig:phase_request_timelines_dsv4_pro}
    \end{subfigure}
    \vspace{-10pt}
    \caption{In-flight request counts for the three pipeline phases (generation, validation, profiling) over the first 10{,}000~seconds of execution, with 10 agent workflows and 10 exclusive-mode validation/profiling GPUs.}
    \label{fig:phase_request_timelines}
    \vspace{-15pt}
\end{figure*}
\subsection{Insufficient Profiling Feedback}\label{subsec:low_feedback}

Profiling feedback refers to the NCU profiling results of a valid kernel, which the agent uses to guide the next iteration's optimization. 
However, Figure~\ref{fig:iteration_status_statistics} shows that 36.3\% of iterations under GLM-5.1 and 40.7\% under DeepSeek-V4-Pro produce kernels that pass both validation and profiling. 
On seven of the ten tasks, fewer than half the iterations yield profiling feedback.
Failure modes vary by model. GLM fails predominantly at compile time on the harder
matmul variants, indicating brittle code shape and syntax (e.g., templates, kernel signatures, includes).
DeepSeek shows more runtime failures, suggesting it compiles more aggressive variants whose indexing,
memory access, or launch configurations are unsafe. Numerical mismatches recur on transpose and
triangular matmul for both models, pointing to subtle layout or accumulator bugs that pass static checks but
produce incorrect outputs.

\begin{table}[t]
    \centering
    \scriptsize
    \setlength{\tabcolsep}{2pt}
    \caption{Best speedups over the \texttt{KernelBench} reference from 100 non-reasoning generations w/o and w/ conditioning on \emph{reasoning prefixes} with two models.
\ding{55}: no available kernels.}
    \label{tab:nonreasoning_speedups_w_wo_shared_prefix_cache}
    \vspace{-10pt}
    {
        \renewcommand{\arraystretch}{0.85}
        \resizebox{\linewidth}{!}{
            \begin{tabular}{lcccccccccc}
                \toprule
                Method & T1 & T2 & T3 & T4 & T5 & T6 & T7 & T8 & T9 & T10 \\
                \midrule
                GLM w/o      & \ding{55} & 0.44 & \ding{55} & \ding{55} & \ding{55} & \ding{55} & \ding{55} & \ding{55} & \ding{55} & 1.00 \\
                GLM w/       & 8.54 & 2.79 & 0.79 & 58.16 & 4.21 & 3.41 & 2.90 & 4.69 & 4.00 & 4.25 \\
                \midrule
                DSv4 w/o & \ding{55} & \ding{55} & \ding{55} & 57.27 & 0.36 & 0.81 & 0.35 & 0.44 & \ding{55} & 0.02 \\
                DSv4 w/  & 8.76 & 0.68 & 0.88 & 61.32 & 4.04 & 6.35 & 2.20 & 3.51 & 1.33 & 0.73 \\
                \bottomrule
            \end{tabular}
        }
    }
    \vspace{-15pt}
\end{table}
\noindent\textbf{Reason.} The insufficient profiling feedback stems from a gating effect between validation and profiling. Profiling measures real performance metrics on the GPU using NCU, and is meaningless for incorrect kernels. The agent therefore only profiles kernels that pass validation. The remaining iterations fail validation, never reaching the profiler.


\noindent\textbf{Implication 2.} The insufficient profiling feedback forces much of the budget into failed candidates rather than
performance-improving exploration. This motivates mechanisms that increase the number of valid candidates reaching
profiling.

\subsection{Underutilized Validation/Profiling Resources}\label{subsec:gpu_idle}

Under the ``one GPU per kernel'' partitioning introduced in \S\ref{sec:bg}, the validation/profiling resource pool is severely underutilized.
Figure~\ref{fig:phase_request_timelines} shows the in-flight request counts for the three phases over the first 10{,}000 seconds of agentic kernel optimization.
We observe that the validation/profiling resource pool sustains only 0.14--1.68 concurrent requests per second on average, one to two orders of magnitude below the 8.19--9.62 concurrent generation requests over the same time window.
The resource utilization is 4.7\% on GLM and 11.3\% on DeepSeek.
The validation and profiling load is thus one to two orders of magnitude lower than generation.

\noindent\textbf{Reason.}
First, candidate kernels emerge only after the generation completes, so validation
GPUs remain idle during most of the generation phase.
Second, many candidates fail validation and never reach profiling, leaving
profiling GPUs with even fewer requests.

\noindent\textbf{Implication 3.} Validation and profiling GPUs sit chronically idle during the generation phase, 
leaving GPU resources underutilized. This motivates mechanisms that make candidate kernels available while reasoning
generation is still running.

\begin{figure}[t]
    \centering
    \includegraphics[width=\linewidth]{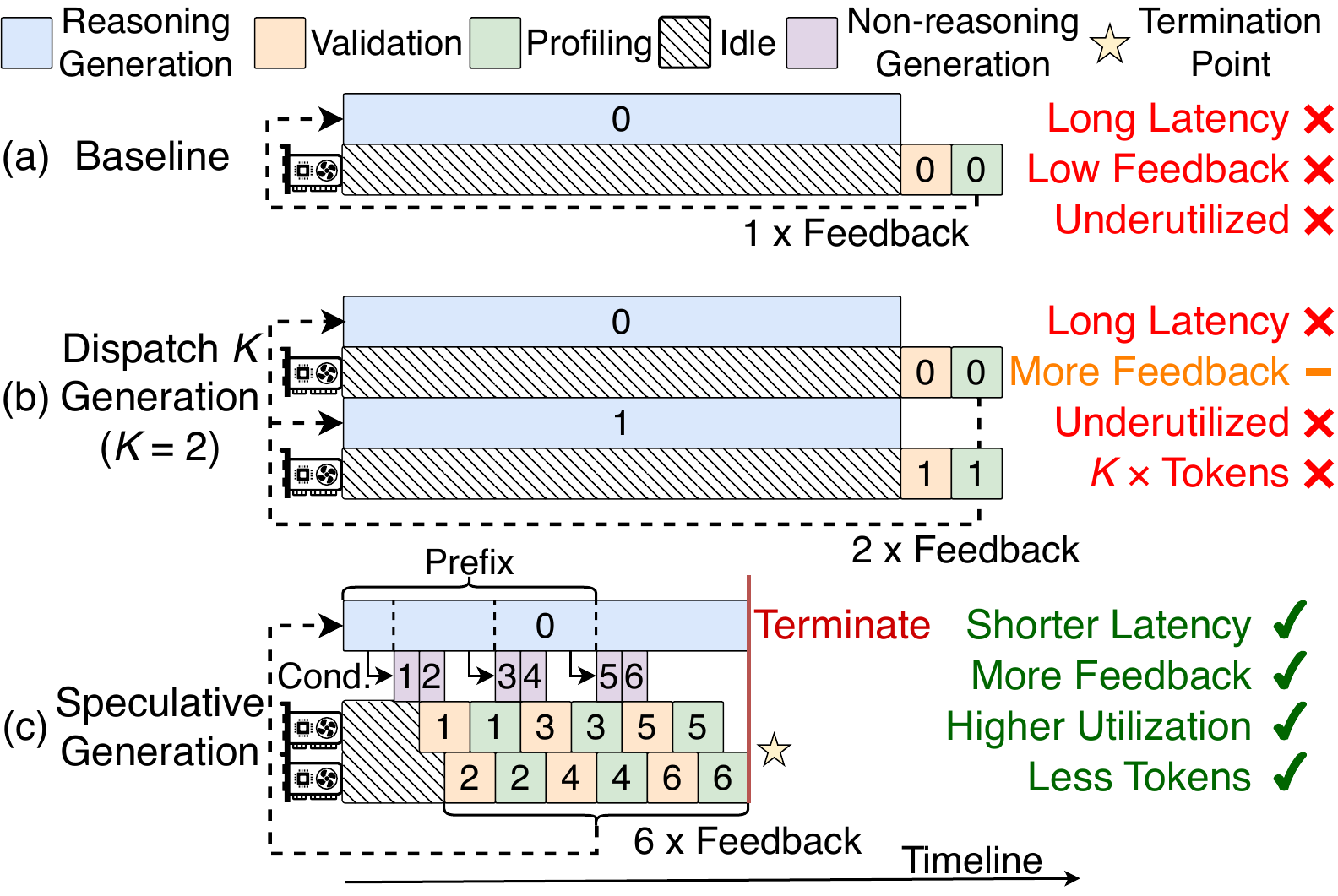}
    \vspace{-20pt}
    \caption{Comparison of existing techniques in agentic kernel optimization and speculative generation.}
    \vspace{-15pt}
    \label{fig:timeline_workflow}
\end{figure}

\section{Design Insight and Challenges}\label{sec:challenges}

\subsection{Limitations of Prior Systems}\label{subsec:limitations_of_prior_systems}

As illustrated in \S\ref{sec:mot}, the baseline iterative refinement framework~\cite{kernelbench}, which performs one generation per iteration (see Figure~\ref{fig:timeline_workflow}(a)),
suffers from three system-level inefficiencies.
An intuitive approach is to dispatch multiple generations per iteration to improve search efficiency.
However, (1) dispatching multiple \emph{reasoning generations}~\cite{cudaforge,K-Search,kernelagent2025,AlphaEvolve,Kernel-Smith} 
roughly multiplies the token cost by $K$ (see Figure~\ref{fig:timeline_workflow}(b)).
The candidate kernels become actionable only after the reasoning process finishes, 
resulting in significant idle time for validation/profiling resources.
(2) Dispatching multiple \emph{non-reasoning generations} reduces both generation latency and token cost but rarely produces valid or performant kernels.
Table~\ref{tab:nonreasoning_speedups_w_wo_shared_prefix_cache} shows that 8/10 tasks on GLM 
and 4/10 tasks on DeepSeek fail to produce any valid kernels after 100 iterations of non-reasoning generations.

\subsection{Insight: Speculative Generation}\label{subsec:design_insight}
Our key insight is that the reasoning process exposes a window for producing
additional candidate kernels with \emph{reasoning prefix conditioning}.
Since non-reasoning generations are faster than reasoning generations, the system can dispatch
multiple non-reasoning generations in parallel with the ongoing reasoning generation to 
increase the candidate kernel count.
To improve the kernel performance of non-reasoning generations, we condition non-reasoning generations
on the prefix of the reasoning output. Table~\ref{tab:nonreasoning_speedups_w_wo_shared_prefix_cache} shows 
that with this conditioning, non-reasoning generations produce valid kernels on all
tasks and achieve higher speedups. Moreover, because this conditioning shares the reasoning prefix with the reasoning
generation, speculative generations can reuse the prefix cache and reduce token
cost.

\begin{figure}[t]
    \centering
    \includegraphics[width=\linewidth]{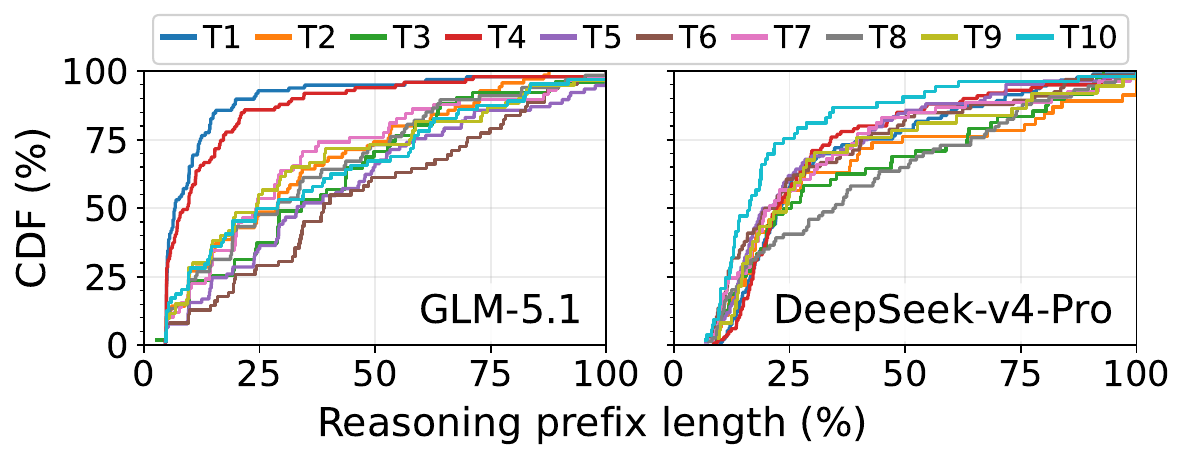}
    \vspace{-25pt}
    \caption{CDF of the shortest reasoning prefix length for generating a kernel with higher speedups than the average speedup of previously generated kernels across both models.}
    \vspace{-10pt}
    \label{fig:spec_reach95_token_cdf}
\end{figure}
We further explore early termination opportunities with prefix conditioning in Figure~\ref{fig:spec_reach95_token_cdf}.
We record the shortest reasoning prefix that produces a kernel faster than the average of previously generated kernels.
Across GLM-5.1 and DeepSeek-V4-Pro, the prefix length ranges from 14\% to 70\% of the full reasoning trace.
These results show that prefix conditioning can provide useful context for non-reasoning generations 
to produce performant kernels before the full reasoning trace completes.
Once a satisfactory kernel appears,
the system can terminate the reasoning generation early to reduce generation latency. We empirically show that a 
pragmatic termination criterion can reduce generation latency without sacrificing kernel performance 
(see \S\ref{subsec:eval_termination_analysis}).

This insight motivates an iteration-level speculative generation scheme.
Compared with the prior approaches in Figure~\ref{fig:timeline_workflow}(b),
speculative generation in Figure~\ref{fig:timeline_workflow}(c) introduces three benefits:
\textbf{(1)} shorter generation latency by enabling early termination when a
forked kernel meets the termination criterion
\textbf{(C1)};
\textbf{(2)} more profiling feedback via generating more candidate kernels per
iteration
\textbf{(C2)};
\textbf{(3)} higher validation/profiling utilization by keeping
resources busy during the generation phase
\textbf{(C3)}.

\subsection{Challenges of Speculative Generation}\label{subsec:challenges_list}

Realizing speculative generation poses two challenges.

\noindent\textbf{Challenge 1. How to fork and when to terminate.}
Reasoning traces contain noise, such as self-talk and repetition, that disturb
the parsing of useful context for non-reasoning generations.
The system must identify useful conditioning context from a noisy reasoning
trace and decide how many speculative generations to launch.
Too few forks leave validation/profiling resources idle, while too many overload
the queues and delay profiling feedback.
The system should also carefully choose the termination criterion.
Terminating too early sacrifices kernel performance, while
terminating too late saves negligible generation time.


\noindent\textbf{Challenge 2. How to manage bursty speculative generation load.}
Speculative generation reshapes the validation and profiling load from sporadic to bursty.
The legacy static ``one GPU per kernel'' partitioning cannot handle this burstiness.
It leaves some GPUs idle while others queue, so resource allocation must adapt as bursts arrive.
Bounding the profiling feedback latency also requires request prioritization.
For local LLM deployments, bursty speculative generations further introduce
additional memory overhead that the system needs to manage.

\section{System Overview}\label{sec:overview}

\noindent\textbf{Overview.} We introduce \sysname{}, an agentic kernel
optimization system comprising \emph{SpecController} and
\emph{ElasticScheduler}.
SpecController utilizes the ongoing reasoning generation as a window for speculative
generation.
It forks non-reasoning generations conditioned on the reasoning prefix while
keeping the reasoning generation as a fallback.
Once a satisfactory kernel appears, SpecController terminates the
reasoning generation.
ElasticScheduler dynamically reallocates the validation/profiling resource pool,
prioritizes validation and profiling requests, and uses spare memory on
validation/profiling GPUs as remote KV-cache storage for reasoning prefixes.
\sysname{} requires \emph{no changes} to the underlying LLM or search algorithm.
It reduces generation latency, produces more profiling feedback, and keeps
validation/profiling resources busy.

\begin{figure}[t]
    \centering
    \includegraphics[width=\linewidth]{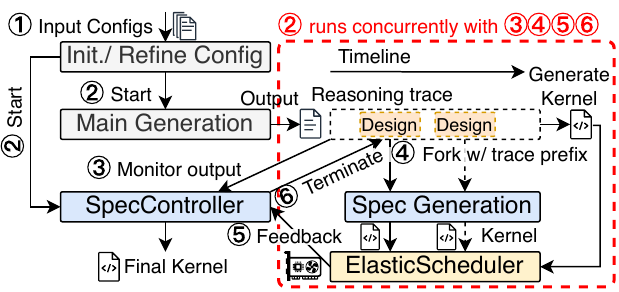}
    \vspace{-20pt}
    \caption{System overview of \sysname{}.}
    \label{fig:system_overview}
    \vspace{-15pt}
\end{figure}

\noindent\textbf{Workflow.}\label{subsec:workflow}
One iteration of \sysname{} proceeds in the six steps below.
\noindent\textbf{(1) Input configuration.}
The user supplies three inputs: the reasoning LLM that will drive optimization,
the prompt template that describes the target kernel and feedback format, and
the search algorithm
(e.g., evolutionary search~\cite{AlphaEvolve,kernelagent2025,openevolve},
best-of-$N$~\cite{K-Search,cudaforge,TritonForge})
that governs how optimization status is updated between iterations.
Optionally, the user supplies a termination criterion for early stopping.
\noindent\textbf{(2) Start main generation and SpecController.}
\sysname{} starts the main reasoning generation.
At the same time, SpecController attaches to the main generation's reasoning output stream as a monitor.
The main generation remains live while steps~(3)--(6) execute, so these steps run concurrently with step~(2).
\noindent\textbf{(3) Monitor output.}
SpecController monitors the main generation's reasoning trace and detects
when the trace has committed to concrete kernel design, such as a tile-shape
choice, a parallelization plan, or a fenced CUDA code block.
These kernel-design decisions serve as trigger signals for non-reasoning
speculative generations.
\noindent\textbf{(4) Fork speculative generation.}
When SpecController emits a trigger signal, it concatenates the iteration's prompt
with the reasoning trace prefix to form a speculative prompt.
It forks speculative generations with this speculative prompt.
As each fork emits a candidate kernel, SpecController dispatches the kernel to
ElasticScheduler for validation and profiling.
SpecController maintains a history of kernels and their speedups to evaluate
the termination criterion.
\noindent\textbf{(5) Elastic scheduling for validation and profiling.}
ElasticScheduler receives validation/profiling requests from SpecController
and returns validation/profiling feedback to SpecController.
It dynamically partitions the GPU pool between validation and profiling to
absorb bursty arrivals, keep resources busy, and bound profiling feedback
latency.
\noindent\textbf{(6) Termination and output final kernel.}
SpecController analyzes the validation/profiling feedback to decide when to terminate the main reasoning.
If a speculative kernel meets the termination criterion, SpecController terminates the main generation and outputs this kernel as the final result.
If no early termination fires, the main reasoning generation remains a fallback,
and \sysname{} selects the best kernel found in the iteration.

\section{Methodology}\label{sec:design}

This section details the two components of \sysname{}: \emph{SpecController} (\S\ref{subsec:speclauncher}), which decides when and how to fork speculative generations, and \emph{ElasticScheduler} (\S\ref{subsec:asyncscheduler}), which dynamically resizes the resource pool and orchestrates concurrent validation and profiling requests.

\begin{figure}[t]
    \centering
    \includegraphics[width=\linewidth]{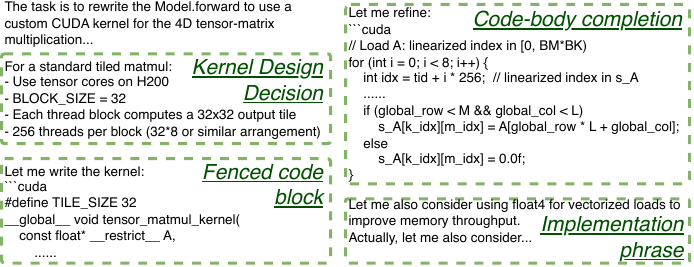}
    \vspace{-20pt}
    \caption{A reasoning trace of GLM-5.1 on the 4D tensor Matmul task. Trigger signals are highlighted in red.}
    \label{fig:kerneldesignintrace}
    \vspace{-15pt}
\end{figure}
\subsection{SpecController}\label{subsec:speclauncher}

Algorithm~\ref{alg:speclauncher} shows when SpecController forks speculative generations and when it 
terminates the reasoning generation.
SpecController monitors the reasoning generation's output. At selected trigger points, 
it concatenates this iteration's prompt with the reasoning prefix to form a speculative prompt.
Conditioned on this speculative prompt, SpecController forks non-reasoning generations 
and dispatches their emitted kernels to validation and profiling while the reasoning generation 
continues as a fallback.
Once a kernel meets the termination criterion, SpecController terminates the reasoning generation
and all pending speculative generations to reduce generation latency.
We next detail when SpecController triggers speculative generations~(\S\ref{subsubsec:dynparser}) and when it terminates the reasoning generation~(\S\ref{subsubsec:dispatch}).

\begin{algorithm}[t]
\caption{SpecController core loop.}
\label{alg:speclauncher}
\begin{algorithmic}[1]
\Require Reasoning LLM $M$, prompt $P_0$, search algorithm $\mathcal{A}$, resource capacity $C$, \#iterations $N$
\State $H \gets \{0\}$;\, $P \gets P_0$ \Comment{Speedup history; initial prompt}
\State $k^{\star} \gets \bot$ \Comment{Best kernel found so far}
\For{$i = 1$ \textbf{to} $N$}
    \State $et \gets \textbf{false}$ \Comment{Early termination flag}
    \State $\mathcal{G} \gets$ stream of $M(P)$ \Comment{Reasoning generation}
    \While{$\mathcal{G}$ is not terminated}
        \If{$\mathcal{G}$ emits a trigger signal \textbf{or} GPU is idle}
            \State $\pi \gets$ prefix of $\mathcal{G}$
            \State $P_s \gets CONCAT(P, \pi)$ \Comment{Speculative prompt}
            \State $K \gets \max(1, \min(C.\textit{val}, C.\textit{prof}))$ \label{line:num_spec_gens}
            \State Fork $K$ spec. generations with prompt $P_s$.
        \EndIf
        \For{each spec. generation $G_s$ with kernel $k_s$}
            \State $speedup_s \gets$ measured speedup of $k_s$
            \State $\tau \gets \mathrm{mean}(H)$
            \State $H \gets H \cup \{speedup_s\}$
            \If{$speedup_s > \tau$}
                \State $k^{\star} \gets k_s$; \, $et \gets \textbf{true}$
                \State Terminate $\mathcal{G}$ and all pending forks;
                \State \textbf{break}
            \EndIf
        \EndFor
    \EndWhile
    \If{$et$ is \textbf{false}}
        \State Update $k^{\star}$ with the best found kernel.
    \EndIf
    \State $P \gets \mathcal{A}(k^{\star},\, H)$ \Comment{Update prompt for next iteration}
\EndFor

\State \Return $k^{\star}$ \Comment{The best found kernel}

\end{algorithmic}
\end{algorithm}

\subsubsection{Trigger Speculative Generation}\label{subsubsec:dynparser}
SpecController uses kernel-design decisions in the reasoning trace as trigger
signals for speculative generations.
Figure~\ref{fig:kerneldesignintrace} shows an example of the reasoning trace from GLM-5.1 on the 4D tensor Matmul task.
During the reasoning process, we observe that the LLM commits concrete
kernel-design decisions.
As reasoning progresses, the LLM implements these decisions in fenced code
blocks or completed kernel bodies.
Finally, the LLM generates the final code based on the reasoning results.
Based on this observation, SpecController uses an empirical parser to detect
these trigger signals.
The parser recognizes four classes of trigger signals.
(1) \emph{Kernel-design decisions.} Concrete kernel-design decisions, such as tile shapes, tile sizes, and specific instructions.
(2) \emph{Fenced code blocks.} Complete code blocks fenced with language tags such as \texttt{cuda}, \texttt{cpp}, or \texttt{python}.
(3) \emph{Kernel-body completion.} A partial kernel body, such as a \texttt{\_\_global\_\_} function with a complete
signature and brace-balanced body.
(4) \emph{Implementation phrases.} Natural-language phrases that announce implementation, such as ``Let me implement\ldots'', ``Here is the plan\ldots''.
We derive these trigger structures from 38{,}745 reasoning traces generated by
GLM-5.1 and DeepSeek-V4-Pro, and implement the parser with regular expressions.

Once SpecController detects these trigger signals, 
it concatenates the iteration's original prompt and the reasoning trace
prefix to construct a speculative prompt.
Then SpecController forks non-reasoning generations with this speculative prompt.
To avoid speculative-generation starvation when no trigger is detected,
SpecController also forks from the current reasoning prefix whenever
validation/profiling resources become idle.
Each fork event launches $K$ speculative generations, where $K$ is determined by 
the resource capacity $C$ (Algorithm~\ref{alg:speclauncher}, Line~\ref{line:num_spec_gens}).
This lets SpecController launch enough speculative generations to keep resources
busy without queuing too many requests and wasting tokens.

\subsubsection{Early Termination}\label{subsubsec:dispatch}
SpecController terminates the reasoning generation once a speculative
kernel meets the predefined termination criterion.
After a speculative generation emits a candidate kernel, SpecController sends it
to ElasticScheduler (\S\ref{subsec:asyncscheduler}) for validation and profiling.
For each valid kernel, ElasticScheduler returns the measured speedup over the
\texttt{KernelBench} reference.
SpecController maintains a history $H$ of previously profiled kernel speedups
and compares each new kernel's speedup against this history.

\begin{figure}[t]
    \centering
    \includegraphics[width=\linewidth]{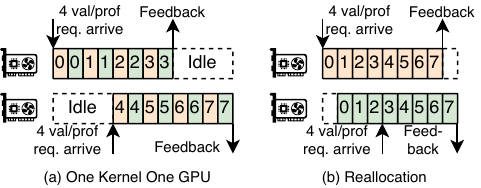}
    \vspace{-20pt}
    \caption{Reducing GPU idle with reallocation.}
    \label{fig:asyncscheduler_workflow}
    \vspace{-15pt}
\end{figure}
In our experiments, we use the mean speedup of previously profiled kernels, $\mathrm{mean}(H)$, as
the termination threshold. When a speculative kernel exceeds this threshold, 
SpecController terminates the reasoning generation and cancels all pending
speculative generations. Otherwise, the reasoning generation continues as a fallback.
This adaptive threshold maintains termination frequency without sacrificing kernel performance.
Figure~\ref{fig:token_kernel_speedups_GLM_5_1} empirically confirms the efficiency of this termination criterion.
As optimization progresses, $\mathrm{mean}(H)$ rises with the performance of previously found kernels.
At the same time, it is less sensitive than $\max(H)$ to a 
single outlier, which could otherwise suppress early termination for many later iterations.
For example, on the Diagonal Matmul task, a kernel reaches
56$\times$ speedup over the reference in iteration 5 and remains the best kernel until iteration 72 (Figure~\ref{fig:token_kernel_speedups_GLM_5_1}).
If the threshold were $\max(H)$, early termination would rarely fire during this
interval, losing generation-latency savings without improving the best kernel.
Instead, using $\mathrm{mean}(H)$ as the threshold allows SpecController to
reduce generation latency through more frequent early termination
(Figure~\ref{fig:e2e_execution_time}).
\sysname{} also exposes interfaces for user-supplied termination criteria.

\algnewcommand\algorithmicasync{\textbf{async}}
\algdef{SE}[ASYNC]{Async}{EndAsync}
{\algorithmicasync}
{\algorithmicend\ \algorithmicasync}

\begin{algorithm}[t]
\caption{ElasticScheduler core loop.}
\label{alg:asyncscheduler}
\begin{algorithmic}[1]
\Require Total GPU pool $G$
\State $Q_v \gets \emptyset$;\, $Q_p \gets \emptyset$ \Comment{Request queues}
\State $L_v \gets 0$;\, $L_p \gets 0$ \Comment{Max queue lengths}
  \For{each iteration $i$}
      \State $(G_{val}, G_{prof}) \gets \Call{Allocate}{L_v, L_p, G}$ \Comment{Reallocation}
      \State $L_v \gets 0$;\, $L_p \gets 0$ \Comment{Reset max queue lengths}
      \Async
      \State Queue requests to $Q_v$ (LAF) and $Q_p$ (FIFO).
      \State $L_v \gets \max(L_v, |Q_v|)$;\, $L_p \gets \max(L_p, |Q_p|)$
      \State Serve requests with $G_{val}$ and $G_{prof}$ on idle GPUs.
      \EndAsync
      \State Terminate in-flight requests and clear $Q_v$ and $Q_p$.
  \EndFor
\end{algorithmic}
\end{algorithm}
\subsection{ElasticScheduler}\label{subsec:asyncscheduler}

Algorithm~\ref{alg:asyncscheduler} illustrates the core loop of ElasticScheduler.
At the beginning of each iteration, ElasticScheduler calls \Call{Allocate}{} to reallocate the validation/profiling GPU split
$(G_v, G_p)$ based on the maximum validation and profiling queue lengths, $L_v$ and $L_p$, from the previous iteration.
ElasticScheduler receives validation and profiling requests from SpecController and serves them on the corresponding GPU split with two priority queues.
At the iteration boundary, SpecController consumes only the validation/profiling
results that have already returned.
ElasticScheduler then aborts remaining requests from the finished iteration and
clears both queues, so speculative tails never delay the next iteration.
Finally, ElasticScheduler repurposes unused GPU memory in the validation/profiling
GPUs as remote KV-cache storage for the reasoning prefix.
We next detail the resource reallocation (\S\ref{subsubsec:priqueue}), the priority queues (\S\ref{subsubsec:reduce_prof_latency}), and the remote
cache of reasoning prefixes (\S\ref{subsubsec:remote_cache}).

\subsubsection{Resource Reallocation}\label{subsubsec:priqueue}

ElasticScheduler divides the total GPU pool into two splits: validation GPUs and profiling GPUs.
The number of GPUs in each split is determined by the max queue length $L_v$ and $L_p$ from the previous iteration.
Specifically, when $L_v + L_p = 0$, ElasticScheduler uses an even split.
Otherwise, it sets $G_{prof} = \min(G-1, \max(1, \left\lceil G \cdot \frac{L_p}{L_v + L_p} \right\rceil)), 
\quad G_{val} = G - G_{prof}$.
With this reallocation, ElasticScheduler can absorb bursty speculative-generation load and reduce GPU idle.
Figure~\ref{fig:asyncscheduler_workflow} shows an example of the resource reallocation.
In this example, the static one-GPU-per-phase allocation keeps one GPU assigned
to validation and one GPU assigned to profiling.
Because validation and profiling requests arrive at different times, one split
can queue requests while the other split remains idle.
ElasticScheduler reallocates the two GPUs according to the previous iteration's
queue pressure, allowing idle GPUs to move toward the heavier-loaded phase.

\subsubsection{Reducing Profiling Feedback Latency}\label{subsubsec:reduce_prof_latency}

ElasticScheduler prioritizes latest-arrival validation requests (LAF) and
earliest-arrival profiling requests (FIFO).
Later validation requests are produced from longer reasoning prefixes.
They are therefore conditioned on more design information than earlier requests.
Later candidates are therefore more likely to contain complete design
information and to meet the early-termination threshold.
LAF intentionally favors fresher candidates within an iteration.
Older validation requests may be skipped at the iteration boundary, but this does
not block progress because the reasoning generation remains a fallback and
the next iteration starts with a cleared queue.
For profiling requests, FIFO avoids delaying an already validated kernel behind
later arrivals.
Because every profiled kernel provides usable feedback, serving the oldest
validated kernel first reduces feedback latency without favoring speculative
freshness.

\subsubsection{Remote Cache of Reasoning Prefixes}\label{subsubsec:remote_cache}

Speculative generations share the reasoning prefix with the main generation.
However, keeping all prefix KV caches in local serving memory can exceed the
memory budget when multiple speculative generations are active.
ElasticScheduler therefore uses spare memory on validation/profiling GPUs as
remote KV-cache storage.
Validation and profiling consume little GPU memory in our workload, even under
bursty speculative load, so the remote cache uses otherwise idle memory without
reducing the memory available to validation or profiling execution.

When local serving memory approaches its capacity limit, ElasticScheduler
migrates KV caches of suspended reasoning prefixes to the remote GPU pool
instead of discarding them.
When a speculative generation later resumes from the same prefix, it restores
the cached KV state rather than recomputing the prefix from tokens~\cite{vLLM}.
The migration uses the Mooncake~\cite{Mooncake} backend for high-throughput device-to-device
RDMA transfers, avoiding the CPU/TCP data path.
This reduces prefix recomputation under a limited memory budget, further improving the system efficiency (see \S\ref{subsec:eval_performance_breakdown}).

\section{Implementation}\label{subsec:impl}

\sysname{} is implemented in approximately 9k lines of Python code.
SpecController monitors the LLM's reasoning trace through the OpenAI API v2.16.0.
SpecController parses the reasoning trace to detect trigger signals with a regular expression parser built on 38{,}745 reasoning traces from GLM-5.1 and DeepSeek-V4-Pro.
ElasticScheduler runs as a microservice to manage the validation and profiling requests.
Validation workers compile each candidate kernel via \texttt{nvcc} driven by Ninja 1.13.0 with \texttt{MAX\_JOBS=20}, then run a single correctness check against the \texttt{KernelBench} reference on the validation GPU.
Profiling workers invoke NCU 2025.1.0 for hardware-counter collection. Both validation and profiling workers run exclusively on GPUs.
The system exposes interfaces for user-defined search algorithms, prompt templates, and termination criteria.
\section{Evaluation}\label{sec:eval}

Our evaluation answers eight questions.
\textbf{Q1 (\S\ref{subsec:eval_execution_time}):} How much does \sysname{} reduce E2E execution time compared with other systems?
\textbf{Q2 (\S\ref{subsec:eval_success_rate}):} How much does \sysname{} increase profiling feedback?
\textbf{Q3 (\S\ref{subsec:eval_resource_utilization}):} Does \sysname{} keep validation/profiling GPUs busy during the generation phase?
\textbf{Q4 (\S\ref{subsec:eval_performance_breakdown}):} Which components contribute to the E2E speedup?
\textbf{Q5 (\S\ref{subsec:eval_kernel_perf}):} Does speculative generation improve final kernel performance rather than trading performance for shorter iterations?
\textbf{Q6 (\S\ref{subsec:eval_token_consumption}):} What additional token overhead does \sysname{} introduce?
\textbf{Q7 (\S\ref{subsec:eval_harder_tasks}):} Does \sysname{} remain effective on harder \texttt{KernelBench} Level~2/3 tasks?
\textbf{Q8 (\S\ref{subsec:eval_termination_analysis}):} How do E2E time and final kernel performance change with different termination criteria?

\begin{table}[t]
    \centering
    \caption{Short names of ten \texttt{KernelBench} Level~2/3 kernel
    optimization tasks in our extended experiments.}
    \vspace{-10pt}
    \label{tab:tasks_level_2_and_3}
    \small
    \setlength{\tabcolsep}{3pt}
    \renewcommand{\arraystretch}{0.9}
    \begin{tabular}{@{}llll@{}}
    \toprule
    \textbf{Abbr.} & \textbf{Task} & \textbf{Abbr.} & \textbf{Task} \\
    \midrule
    T11 & Gemm $\times$ LeakyReLU   & T16 & Conv2d-BN-Scale \\
    T12 & Gemm Div-Sum-Scale         & T17 & Gemm-Add-ReLU \\
    T13 & Gemm-Scale-BN              & T18 & Matmul-GELU-Softmax \\
    T14 & Conv2d-Act-BN              & T19 & MLP (Level 3) \\
    T15 & Matmul-Sigmoid-Sum         & T20 & ReLU Self-Attn (Level 3) \\
    \bottomrule
    \end{tabular}
    \vspace{-15pt}
\end{table}
\subsection{Experimental Setup}\label{subsec:eval_setup}
\noindent\textbf{Testbed \& Models.}
Our experiments use up to 18 H200 GPUs in our internal cluster. Each node is interconnected via NVLink and RoCEv2 and hosts an Intel Xeon Platinum 8558 CPU.
We serve GLM-5.1 via vLLM with 8 GPUs, while the remaining GPUs are dedicated to validation and profiling.
DeepSeek-V4-Pro is accessed via its official API with high reasoning effort.
The software stack includes PyTorch 2.10, CUDA 12.9, NCU 2025.1.0, and Ninja.
For both models, temperature is set to 0.1 to maintain sharp code-generation formats, and speculative forks disable the reasoning traces.

\begin{figure*}[t]
    \centering
    \includegraphics[width=\linewidth]{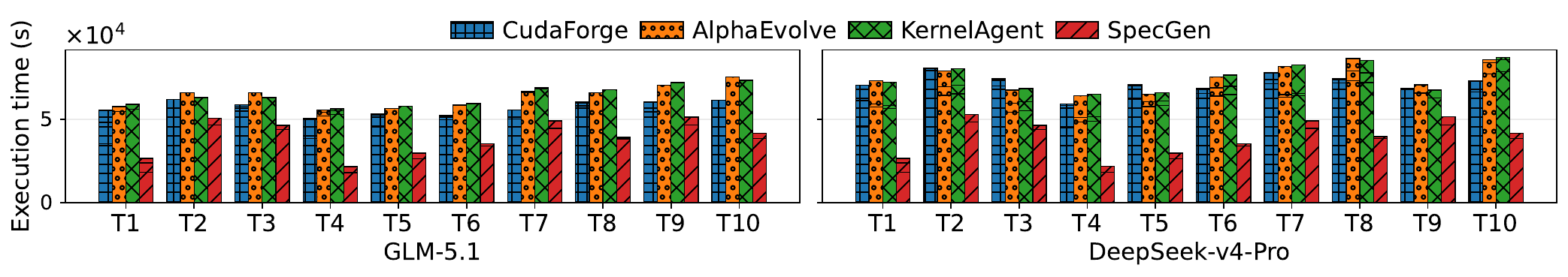}
    \vspace{-27pt}
    \caption{End-to-end execution time of 100 iterations per task on GLM-5.1 and DeepSeek-V4-Pro.}
    \vspace{-15pt}
    \label{fig:e2e_execution_time}
\end{figure*}

\begin{figure*}[t]
    \centering
    \includegraphics[width=\linewidth]{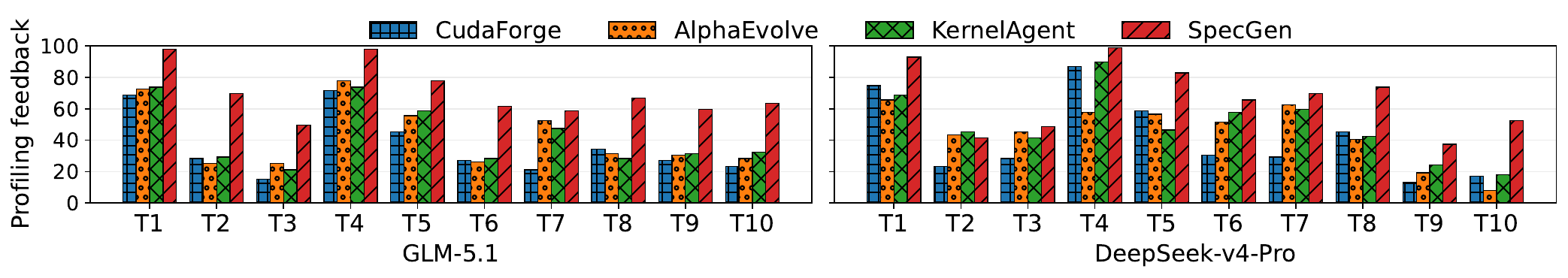}
    \vspace{-26pt}
    \caption{Profiling feedback of 100 iterations per task on GLM-5.1 and DeepSeek-V4-Pro.}
    \vspace{-15pt}
    \label{fig:profiling_feedback_rate_lift}
\end{figure*}

\noindent\textbf{Workloads \& Metrics.}
We evaluate the ten KernelBench Level~1 tasks (Table~\ref{tab:tasks}) and ten Level~2/3 tasks (Table~\ref{tab:tasks_level_2_and_3}) with a budget of 100 iterations per task.
To filter device noise, each kernel's speedup over the PyTorch reference is averaged across 40 timed runs (following 10 warmups).
All cross-task aggregated results report the geometric mean.

\noindent\textbf{Baseline.}
We compare \sysname{} against three representative agentic kernel 
optimization systems.
\textbf{(1) CudaForge}~\cite{cudaforge} is a training-free multi-agent 
framework that employs a Coder-Judge workflow to iteratively generate 
and optimize CUDA kernels, guided by hardware feedback from NCU metrics.
\textbf{(2) AlphaEvolve}~\cite{AlphaEvolve} is a coding 
agent from Google DeepMind that pairs reasoning LLMs with automated 
evaluators, using an evolutionary algorithm to iteratively discover 
and refine kernel performance.
\textbf{(3) KernelAgent}~\cite{kernelagent2025} is an autonomous 
kernel generation system from Meta. It 
combines problem analysis, parallel LLM-assisted kernel 
generation with numerical verification, and a hardware-guided 
iterative optimization loop to tune kernel performance.

\subsection{End-to-End Execution Time}\label{subsec:eval_execution_time}

\sysname{} consistently reduces E2E execution time and provides final kernels with 
higher speedups than the three agentic kernel optimization baselines (see Figure~\ref{fig:token_kernel_speedups_GLM_5_1}).

\noindent\textbf{Overall time speedup.}
Figure~\ref{fig:e2e_execution_time} shows that across ten \texttt{KernelBench} tasks, \sysname{} 
achieves a geomean time speedup of 1.50$\times$ over CudaForge on GLM-5.1 and 1.88$\times$ on
DeepSeek-V4-Pro.
Against AlphaEvolve, the geomean speedups are 1.68$\times$ on GLM-5.1 and
1.96$\times$ on DeepSeek-V4-Pro.
Against KernelAgent, the geomean speedups are 1.69$\times$ and 1.96$\times$,
respectively.
Across both models, \sysname{} improves end-to-end time by 1.68$\times$ over
CudaForge, 1.81$\times$ over AlphaEvolve, and 1.82$\times$ over KernelAgent in
geomean.

\noindent\textbf{Long generation latency prolongs iteration time.}
CudaForge follows an iterative Coder-Judge workflow that waits for each
reasoning generation to finish before validation and profiling can provide
hardware feedback.
AlphaEvolve uses an evolutionary loop with automated evaluators, but each
program variant still becomes actionable only after the LLM finishes producing
it.
KernelAgent adds static analysis, parallel kernel generation, strict numerical
verification, and hardware-guided optimization, but its validation and profiling
pipeline also depends on completed candidate kernels.
As a result, these systems mainly improve the search policy after candidates
exist, while validation/profiling resources remain idle during long reasoning
generations.

\noindent\textbf{Early termination reduces generation latency.}
\sysname{} changes the timing of candidate availability.
SpecController forks non-reasoning generations from reasoning prefixes while the
reasoning generation continues as a fallback.
When a speculative kernel satisfies the termination criterion, \sysname{}
terminates the reasoning generation early and advances the iteration.
This mechanism directly reduces the longest phase of each iteration.


\subsection{Profiling Feedback}\label{subsec:eval_success_rate}

\sysname{} substantially raises profiling feedback through two effects:
it produces more candidate kernels per iteration, and it conditions
non-reasoning generations on reasoning prefixes to improve kernel performance.

\noindent\textbf{Overall feedback increment.}
As shown in Figure~\ref{fig:profiling_feedback_rate_lift}, on GLM-5.1, \sysname{} raises the average profiling feedback to 70.3,
compared with 36.3 for CudaForge, 42.5 for AlphaEvolve, and 42.4 for
KernelAgent.
This corresponds to geomean lifts of 2.13$\times$, 1.78$\times$, and
1.77$\times$, respectively.
On DeepSeek-V4-Pro, \sysname{} reaches 66.4, compared with 40.7 for
CudaForge, 45.1 for AlphaEvolve, and 49.4 for KernelAgent.
The corresponding geomean lifts are 1.84$\times$, 1.61$\times$, and
1.40$\times$.
Across both models, \sysname{} increases profiling feedback by
1.98$\times$ over CudaForge, 1.69$\times$ over AlphaEvolve, and 1.58$\times$
over KernelAgent in geomean.

\noindent\textbf{More candidates per iteration.}
CudaForge, AlphaEvolve, and KernelAgent differ in search policy, but all three
must wait for completed candidate kernels before validation and profiling can
return feedback.
CudaForge uses hardware-feedback-guided agent iteration, AlphaEvolve uses
evolutionary program refinement, and KernelAgent combines analysis, parallel
generation, verification, and hardware-guided tuning.
These mechanisms improve how candidates are selected and refined after they
exist, but they do not expose validation/profiling to useful kernels during the
reasoning generation.
\sysname{} instead forks speculative non-reasoning generations within the same
iteration, so validation sees more candidate kernels under the same iteration
budget.
This increases the chance that at least one candidate passes validation and
reaches profiling.

\noindent\textbf{Faster speculative candidates.}
\sysname{} conditions each speculative generation on the reasoning prefix, which
passes kernel-design decisions from the reasoning trace into the
non-reasoning prompt.
This prefix conditioning improves the performance of speculative kernels and raises
the probability that each additional candidate reaches profiling.
The combination of more and faster candidates improves profiling
feedback across baselines and models.

\subsection{Validation/Profiling Resource Utilization}\label{subsec:eval_resource_utilization}


\begin{table}[t]
\centering
\caption{Validation/profiling resource utilization. Utilization is the percentage of E2E time during which resources are busy. ES refers to ElasticScheduler. CF: CudaForge, AE: AlphaEvolve, KA: KernelAgent, \textbf{SKG: \sysname{}}.}
\vspace{-10pt}
\label{tab:resource_utilization}
\setlength{\tabcolsep}{4.5pt}
\renewcommand{\arraystretch}{0.9}
\begin{tabular}{lccccc}
    \toprule
    Models & CF & AE & KA & SKG w/o ES & SKG \\
    \midrule
    GLM   & 4.7\% & 5.0\% & 4.2\% & 56.2\% & \textbf{88.2\%} \\
    DSv4  & 11.3\% & 17.6\% & 14.2\% & 74.7\% & \textbf{96.1\%} \\
    \bottomrule
\end{tabular}
\vspace{-15pt}
\end{table}

\begin{figure}[t]
    \centering
    \includegraphics[width=\linewidth]{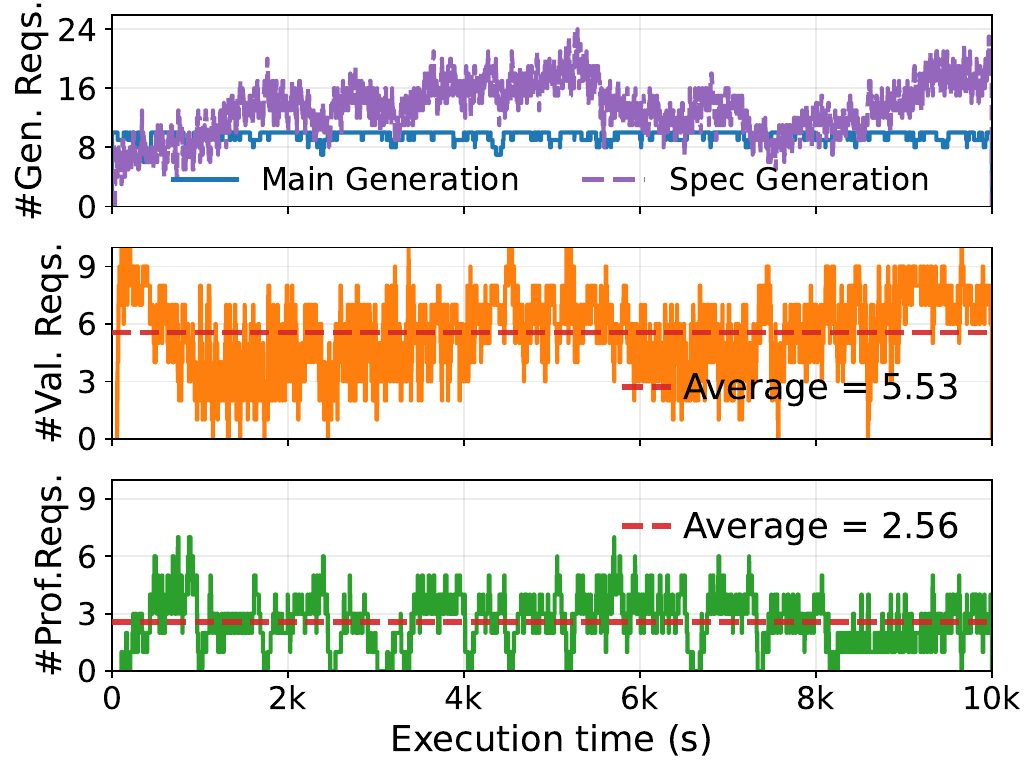}
    \vspace{-25pt}
    \caption{In-flight request counts per phase over the first 10{,}000 seconds of \sysname{} on GLM-5.1.}
    \vspace{-15pt}
    \label{fig:phase_timeline_GLM_5_1}
\end{figure}

\begin{figure*}[t]
    \centering
    \includegraphics[width=\linewidth]{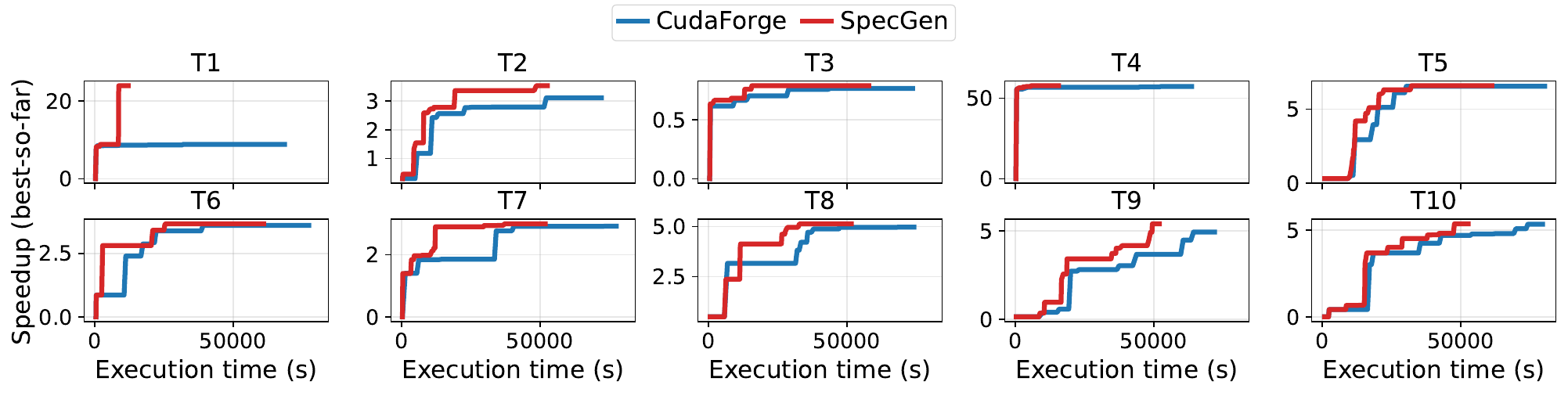}
    \vspace{-25pt}
    \caption{Detailed time-to-speedup over the \texttt{KernelBench} reference across 100 iterations per task on GLM-5.1.}
    \label{fig:token_kernel_speedups_GLM_5_1}
    \vspace{-12pt}
\end{figure*}

\sysname{} substantially improves validation/profiling resource utilization with SpecController and ElasticScheduler.
\noindent\textbf{Overall utilization lift.}
Table~\ref{tab:resource_utilization} shows that the three baseline systems leave
most validation/profiling capacity idle.
On GLM-5.1, their utilization stays between 4.2\% and 5.0\%; on
DeepSeek-V4-Pro, it stays between 11.3\% and 17.6\%.
\sysname{} without ElasticScheduler already raises utilization to 56.2\% on
GLM-5.1 and 74.7\% on DeepSeek-V4-Pro via emitting additional speculative
candidate kernels.
With ElasticScheduler, utilization further rises to 88.2\% and 96.1\%, showing
that SpecController and ElasticScheduler keep the GPU pool busy under bursty arrivals.

\noindent\textbf{Sporadic validation/profiling requests.}
Since long reasoning generations dominate iteration time, validation/profiling
GPUs wait for completed kernels for most of the run.
The feedback bottleneck is amplified because many generated kernels fail
validation and never reach profiling.
Thus, even methods with stronger search logic or parallel candidate generation
can leave the validation/profiling GPU pool idle.

\noindent\textbf{More requests and bursty handling.}
SpecController increases the number of validation/profiling requests by forking
speculative generations during the reasoning generation.
This alone explains the large jump from baseline utilization to \sysname{}
without ElasticScheduler.
However, speculative requests arrive in bursts, which can leave some GPUs busy
while others are idle under a static split.
ElasticScheduler reallocates validation/profiling capacity based on queue
pressure and adjusts the number of speculative forks when validation/profiling GPUs
become idle or busy.
Figure~\ref{fig:phase_timeline_GLM_5_1} illustrates this behavior over the first
10{,}000 seconds, where validation and profiling remain active instead of waiting
for completed reasoning generations.
The result is near-saturated validation/profiling GPU utilization.

\subsection{Performance Breakdown}\label{subsec:eval_performance_breakdown}

\begin{table}[t]
    \centering
    \caption{Incremental performance breakdown of \sysname{} on GLM-5.1 across ten \texttt{KernelBench} tasks. Speedup is geomean end-to-end time speedup over CudaForge.}
    \vspace{-10pt}
    \label{tab:ablation_study}
    \renewcommand{\arraystretch}{0.9}
    \begin{tabular}{lcc}
        \toprule
        Configuration & Speedup & $\Delta$ \\
        \midrule
        Baseline & 1.00$\times$ & 0.00 \\
        + Speculative Generation & 1.46$\times$ & 0.46$\uparrow$ \\
        + Resource Reallocation & 1.61$\times$ & 0.15$\uparrow$ \\
        + Priority Queue & 1.69$\times$ & 0.08$\uparrow$ \\
        + Remote Prefix Cache & 1.77$\times$ & 0.09$\uparrow$ \\
        \bottomrule
    \end{tabular}
    \vspace{-15pt}
\end{table}
Table~\ref{tab:ablation_study} breaks down where \sysname{}'s end-to-end
speedup comes from.
The first step, speculative generation, accounts for most of the gain by raising
geomean speedup from 1.00$\times$ to 1.46$\times$.
Speculative generation shortens the critical path directly. A speculative
kernel can satisfy the termination criterion before the reasoning generation
finishes, so the iteration no longer has to wait for the reasoning generation.

The remaining components make those speculative kernels useful sooner.
Resource reallocation raises speedup to 1.61$\times$ by moving GPUs toward the
busier validation or profiling phase, which reduces idle time under bursty
speculative arrivals.
Priority queues further raise speedup to 1.69$\times$ by validating later-prefix
candidates first and returning profiling feedback in FIFO order.
Together, these scheduling mechanisms reduce validation/profiling waiting time,
which lets SpecController receive results earlier and trigger early termination
sooner.
Remote prefix caching completes the system by raising speedup to 1.77$\times$.
It lets speculative generations reuse the cached reasoning prefix instead of
recomputing shared prompt state, so the system can add speculative branches with
only modest extra generation cost.

\subsection{Kernel Performance}\label{subsec:eval_kernel_perf}

\begin{table}[t]
    \centering
    \small
    \setlength{\tabcolsep}{2pt}
    \caption{Best kernel speedup over the \texttt{KernelBench} reference. The first block reports GLM and the second block reports DeepSeek. \ding{55}: no available kernels. CF: CudaForge, AE: AlphaEvolve, KA: KernelAgent, \textbf{SKG: \sysname{}}.}
    \vspace{-10pt}
    \label{tab:final_kernel_perf}
    \begin{tabular}{lcccccccccc}
        \toprule
        System & T1 & T2 & T3 & T4 & T5 & T6 & T7 & T8 & T9 & T10 \\
        \midrule
        CF   & 8.81 & 3.12 & 0.76 & 57.23 & 6.56 & 3.60 & 2.92 & 4.97 & 4.94 & 5.34 \\
        AE & 12.83 & 1.61 & \ding{55} & 47.43 & 2.45 & 0.81 & 0.69 & 4.15 & 3.98 & 2.36 \\
        KA & 21.23 & 2.21 & \ding{55} & 56.98 & 4.97 & 2.57 & 2.60 & 4.31 & 5.11 & 4.67 \\
        \textbf{SKG}  & \textbf{23.86} & \textbf{3.54} & \textbf{0.79} & \textbf{57.72} & \textbf{6.60} & \textbf{3.66} & \textbf{2.99} & \textbf{5.13} & \textbf{5.41} & \textbf{5.37} \\
        \midrule
        CF   & 8.64 & 0.88 & 0.86 & 60.87 & 4.76 & 3.61 & 2.30 & 3.72 & 0.53 & 0.46 \\
        AE & 7.98 & 0.53 & 0.61 & 59.47 & 1.00 & 3.36 & 1.29 & 3.11 & 0.82 & 0.71 \\
        KA & 6.17 & 0.23 & 0.05 & 61.15 & 4.98 & 4.32 & 2.29 & 3.50 & 1.02 & 0.69 \\
        \textbf{SKG}  & \textbf{8.76} & \textbf{1.69} & \textbf{0.90} & \textbf{61.54} & \textbf{5.38} & \textbf{5.94} & \textbf{3.00} & \textbf{3.87} & \textbf{1.19} & \textbf{0.73} \\
        \bottomrule
    \end{tabular}
    \vspace{-15pt}
\end{table}

\sysname{} improves kernel performance and finds faster kernels within
fixed time budget with early termination.

\noindent\textbf{Overall kernel performance.}
Table~\ref{tab:final_kernel_perf} shows the best kernel speedup over
the \texttt{KernelBench} reference after 100 iterations.
On both GLM-5.1 and DeepSeek-V4-Pro, \sysname{} obtains the best final kernel on
all ten tasks.
Its geomean speedup over the reference reaches 5.78$\times$ on GLM-5.1 and
3.49$\times$ on DeepSeek-V4-Pro.
Across both models, this corresponds to geomean lifts of 1.24$\times$ over
CudaForge, 1.91$\times$ over AlphaEvolve, and 1.52$\times$ over KernelAgent.

\noindent\textbf{Faster kernel within fixed time budget.}
In this experiment, \sysname{} stops the reasoning generation only after a
speculative kernel exceeds the average speedup of previously profiled kernels.
Figure~\ref{fig:token_kernel_speedups_GLM_5_1} shows that with early termination,
\sysname{} finds faster kernels than CudaForge within the fixed time budget.
Across all tasks, the best-so-far speedup trajectory of \sysname{} reaches or exceeds
the CudaForge trajectory by the end of the execution.
The final results in Table~\ref{tab:final_kernel_perf} further confirm that
speculative generation and early termination improve kernel performance while reducing
the end-to-end execution time.

\subsection{Overhead of \sysname{}}\label{subsec:eval_token_consumption}

\sysname{} introduces modest token overhead with prefix caching and early termination.

\begin{table}[t]
    \centering
    \small
    \setlength{\tabcolsep}{2.3pt}
    \caption{Token consumption (millions) of \sysname{} relative to CudaForge across 100 iterations on GLM-5.1. Ratio is \sysname{} tokens divided by CudaForge tokens.}
    \vspace{-10pt}
    \label{tab:token_overhead}
    \begin{tabular}{lcccccccccc}
        \toprule
        & T1 & T2 & T3 & T4 & T5 & T6 & T7 & T8 & T9 & T10 \\
        \midrule
        CudaForge & 1.98 & 2.42 & 2.42 & 2.05 & 2.64 & 2.38 & 2.50 & 2.41 & 2.38 & 2.47 \\
        \sysname{} & 0.89 & 2.71 & 2.51 & 1.06 & 3.13 & 2.72 & 2.51 & 2.55 & 2.44 & 2.62 \\
        Ratio & 0.45 & 1.12 & 1.04 & 0.52 & 1.18 & 1.14 & 1.01 & 1.06 & 1.02 & 1.06 \\
        \bottomrule
    \end{tabular}
    \vspace{-15pt}
\end{table}

\noindent\textbf{Overall token cost.}
Table~\ref{tab:token_overhead} shows that \sysname{}'s total token consumption is
slightly lower than CudaForge despite launching speculative generations.
Across all ten tasks, CudaForge consumes 23.66M tokens, while \sysname{} consumes
23.14M tokens, or 0.98$\times$ of CudaForge.
Most matmul tasks incur modest extra tokens, with ratios between 1.01$\times$
and 1.18$\times$.
However, HingeLoss (T1) and diagonal matmul (T4) terminate early in many
iterations.
Their reasoning generations are therefore cut short, saving most of the
tokens that would otherwise be spent after the useful speculative kernel has
already appeared.
As a result, T1 and T4 consume only 0.45$\times$ and 0.52$\times$ of CudaForge's
tokens, which offsets the modest overhead on harder tasks.

\noindent\textbf{Prefix caching and early termination offset token overhead.}
\sysname{} concatenates the original prompt with the current
reasoning prefix to form the speculative prompt. Therefore, the prefill of speculative generations can reuse the KV cache of the reasoning generation.
Moreover, early termination also saves tokens.
Both effects offset the modest overhead across these ten tasks.

\subsection{Performance on Harder Tasks}\label{subsec:eval_harder_tasks}
\begin{table}[t]
\centering
\caption{Results on T11--T20 (Level~2/3) with DeepSeek. Each task runs 100 iterations. CF: CudaForge, SKG: \textbf{\sysname{}}.}
\vspace{-10pt}
\label{tab:comparison-t11-t20}
\small
\setlength{\tabcolsep}{2.3pt}
\renewcommand{\arraystretch}{0.9}
\begin{tabular}{@{}l cc cc cc cc cc@{}}
    \toprule
    & \multicolumn{2}{c}{E2E (k s)} & \multicolumn{2}{c}{Prof.\ FB} & \multicolumn{2}{c}{Util. (\%)} & \multicolumn{2}{c}{Speedup} & \multicolumn{2}{c}{Tokens (M)} \\
    \cmidrule(lr){2-3}\cmidrule(lr){4-5}\cmidrule(lr){6-7}\cmidrule(lr){8-9}\cmidrule(lr){10-11}
    Task & CF & SKG & CF & SKG & CF & SKG & CF & SKG & CF & SKG \\
    \midrule
    T11 & 37.6 & \textbf{24.5} & 21.2 & \textbf{41.3} & 13.6 & \textbf{94.3} & 0.60 & \textbf{1.25} & 1.20 & 1.27 \\
    T12 & 32.2 & \textbf{21.0} &  1.0 & \textbf{13.2} & 14.9 & \textbf{81.2} & 0.01 & \textbf{0.42} & 0.92 & 1.45 \\
    T13 & 39.3 & \textbf{21.5} & 22.2 & \textbf{63.6} & 14.5 & \textbf{84.8} & 0.63 & \textbf{0.63} & 1.40 & 1.12 \\
    T14 & 34.4 & \textbf{22.6} & 37.4 & \textbf{51.5} & 19.2 & \textbf{91.2} &  1.65 & \textbf{1.68} & 1.01 & 1.22 \\
    T15 & 40.2 & \textbf{18.3} & 29.3 & \textbf{69.7} & 15.2 & \textbf{86.3} &  0.74 & \textbf{0.77} & 1.19 & 1.96 \\
    T16 & 35.8 & \textbf{17.0} & 37.4 & \textbf{73.7} & 19.2 & \textbf{92.4} &  1.06 & \textbf{1.27} & 1.27 & 1.18 \\
    T17 & 38.3 & \textbf{22.6} & 24.2 & \textbf{55.6} & 15.2 & \textbf{88.5} &  0.62 & \textbf{0.74} & 1.14 & 1.55 \\
    T18 & 32.7 & \textbf{27.7} &  7.1 & \textbf{31.3} &  8.5 & \textbf{81.1} &  0.35 & \textbf{55.79} & 1.18 & 1.10 \\
    T19 & 61.5 & \textbf{49.0} & 41.9 & \textbf{69.2} & 22.3 & \textbf{93.2} &  0.64 & \textbf{1.05}  & 2.94 & 3.17 \\
    T20 & 55.1 & \textbf{47.2} & 39.4 & \textbf{77.1} & 14.0 & \textbf{89.2} &  1.10 & \textbf{1.39}  & 2.61 & 2.92 \\
    \bottomrule
\end{tabular}
\vspace{-15pt}
\end{table}

\sysname{} remains effective on harder \texttt{KernelBench} Level~2/3 tasks,
improving runtime, feedback density, validation and profiling resource utilization, and
final kernel speedup over the reference with modest token overhead.
Table~\ref{tab:comparison-t11-t20} compares \sysname{} with CudaForge on ten
harder tasks with DeepSeek.
On average, \sysname{} reduces E2E time by 1.57$\times$ and raises profiling
feedback from 26.1 to 54.6.
It also increases resource utilization from 15.7\% to 88.2\%.
Moreover, \sysname{} finds kernels at least as fast as CudaForge on all ten tasks and a
strictly faster kernel on nine of them, improving geomean kernel speedup over
the reference from 0.49$\times$ to 1.42$\times$.
The additional token cost remains moderate, with a 14.0\%
increase in total token consumption.

\subsection{Termination Criterion Analysis}\label{subsec:eval_termination_analysis}

\sysname{} improves kernel performance within a fixed token budget and
exposes a tunable trade-off between E2E time and kernel performance.

\noindent\textbf{More feedback per token yields faster kernels.}
Table~\ref{tab:termination_analysis} compares \sysname{} against CudaForge under
matched token budgets.
Simply doubling CudaForge's token budget barely improves performance, raising the
final speedup from 5.05$\times$ to 5.15$\times$ on GLM and from 2.61$\times$
to 2.74$\times$ on DeepSeek.
\sysname{} with the historical-average criterion reaches 5.78$\times$ and
3.18$\times$ while consuming only 130.2\% of CudaForge's tokens, fewer than
CudaForge$^+$.
The reason is that speculative generations are non-reasoning and reuse the
cached reasoning prefix, so each candidate costs few tokens.
Under the same budget, \sysname{} therefore produces far more profiling feedback
(114.9 and 101.1 additional feedback) than extra reasoning generations would.
This added feedback in return guides the LLM toward faster kernels.

\noindent\textbf{The termination threshold tunes the time-performance trade-off.}
Lowering the threshold trades kernel performance for shorter runtime.
First-valid termination stops earliest, giving the shortest E2E time, 
yet still emits many candidate kernels and reaches
5.53$\times$ and 2.95$\times$, already faster than CudaForge.
The historical-average default sits in the middle, preserving most of the latency
benefit while improving performance.
When tokens are abundant, disabling termination finds the fastest kernels at the cost of more tokens.
Across every criterion, \sysname{} produces faster kernels than CudaForge under a
comparable or smaller token budget.

\begin{table}[t]
    \centering
    \small
    \setlength{\tabcolsep}{3pt}
    \caption{Comparison of CudaForge and \sysname{} with different termination criteria across T1--T10. The first block reports results with GLM and the second reports DeepSeek. CudaForge$^+$: Doubled token budget for CudaForge.}
    \vspace{-10pt}
    \label{tab:termination_analysis}
    \renewcommand{\arraystretch}{0.9}
    \begin{tabular}{lccccc}
        \toprule
        \makecell[l]{Method} & \makecell[c]{\#Additional\\feedback} & \makecell[c]{Kernel\\Speedup} & \makecell[c]{Token\\Ratio} & \makecell[c]{\#Term.} & \makecell[c]{E2E\\time (s)} \\
        \midrule
        CudaForge & 0 & 5.05$\times$ & 100\% & 0 & 73.9k \\
        CudaForge$^+$  & 35.2 & 5.15$\times$ & 208.6\% & 0 & 75.2k \\
        First valid  & 70.0 & 5.53$\times$ & 114.3\% & 70.3 & 39.8k \\
        \textbf{Hist. avg.}  & 114.9 & 5.78$\times$ & 130.2\% & 63.0 & 43.8k \\
        Hist. best  & 535.9 & 6.06$\times$ & 229.5\% & 9.0 & 70.3k \\
        No term.  & 587.2 & 6.15$\times$ & 240.0\% & 0 & 73.9k \\
        \midrule
        CudaForge  & 0 & 2.61$\times$ & 100\% & 0 & 57.0k \\
        CudaForge$^+$  & 44.3 & 2.74$\times$ & 210.9\% & 0 & 57.7k \\
        First valid  & 64.5 & 2.95$\times$ & 105.1\% & 64.5 & 29.6k \\
        \textbf{Hist. avg.}  & 101.1 & 3.18$\times$ & 130.2\% & 52.8 & 35.9k \\
        Hist. best  & 252.3 & 4.79$\times$ & 208.3\% & 8.1 & 54.6k \\
        No term.  & 269.1 & 5.08$\times$ & 218.1\% & 0 & 57.0k \\
        \bottomrule
    \end{tabular}
    \vspace{-15pt}
\end{table}
\section{Related Work}\label{sec:related}

\noindent\textbf{Training-based kernel optimization.}
Recent work improves CUDA and Triton kernel generation by adapting the \emph{model} itself through reinforcement learning and post-training
over execution feedback.
Kevin~\cite{kevin32b}, CUDA-L1~\cite{cudal1}, and CUDA Agent~\cite{cudaagent} define the basic recipe: optimize the LLM on execution outcomes,
pairwise fast-vs.-slow comparisons, or large-scale agentic RL curricula to raise intrinsic kernel performance.
Later systems broaden this recipe with memory-augmented in-context RL and cross-task reuse~\cite{kernelblaster}, 
hierarchical strategy decomposition~\cite{qimengkernel,tritonrl}, stronger training environments and synthetic 
curricula~\cite{drkernel,drtriton}, and frontier-model fine-tuning~\cite{gpt5kernelgen}.
The common pattern is to spend additional training compute and infrastructure to improve the generator itself.
\sysname{} is complementary: we leave the LLM unchanged and instead reorganize how the agent loop consumes its
outputs, so the gains from better-trained kernel models and from our runtime can compose rather than substitute for one another.

\noindent\textbf{Speculative decoding.}
Speculative decoding~\cite{specdec,specsample} accelerates a \emph{single} LLM generation by drafting several
future tokens and verifying them in parallel. Medusa~\cite{medusa} removes the separate draft model with lightweight
decoding heads, and later variants explore feature-level drafting, self-speculation, adaptive depth, and simpler draft architectures~\cite{eagle,hydra,kangaroo,swiftspecdec,ddd,beagle}.
All of these methods remain strictly \emph{token-level}, speculatively generating tokens to raise throughput.
\sysname{} instead operates at the \emph{iteration-level}, speculatively generating kernels to raise the kernel count per iteration.
The two forms of speculation are orthogonal.
\section{Conclusion}\label{sec:conc}

This paper presents \sysname{}, an efficient agentic kernel optimization system. 
We propose \emph{speculative generation} to address the system-level inefficiencies
of agentic kernel optimization. We design SpecController and ElasticScheduler to manage
the speculative generations and validation/profiling resources, respectively.
Extensive experiments show the effectiveness of \sysname{} in improving the 
agentic kernel optimization efficiency against baseline systems.

\bibliographystyle{plain}
\bibliography{refs}

\end{document}